\documentstyle[11pt]{article}

\begin{document}
\input{psfig}
\large

\def\lsim{\mathrel{\rlap{\lower3pt\hbox{\hskip0pt$\sim$}}
    \raise1pt\hbox{$<$}}}         
\def\gsim{\mathrel{\rlap{\lower4pt\hbox{\hskip1pt$\sim$}}
    \raise1pt\hbox{$>$}}}         
\def\dblint{\mathop{\rlap{\hbox{$\displaystyle\!\int\!\!\!\!\!\int$}}
    \hbox{$\bigcirc$}}}
\def\ut#1{$\underline{\smash{\vphantom{y}\hbox{#1}}}$}

\newcommand{\beq}{\begin{equation}}
\newcommand{\eeq}{\end{equation}}
\newcommand{\dem}{\Delta M_{\mbox{B-M}}}
\newcommand{\dega}{\Delta \Gamma_{\mbox{B-M}}}

\newcommand{\ind}[1]{_{\begin{small}\mbox{#1}\end{small}}}
\newcommand{\WA}{{\em WA}}
\newcommand{\SM}{Standard Model }
\newcommand{\QCD}{{\em QCD }}
\newcommand{\KM}{{\em KM }}
\newcommand{\hscale}{\mu\ind{hadr}}
\newcommand{\sG}{i\sigma G}

\newcommand{\MS}{\overline{\mbox{MS}}}
\newcommand{\pole}{\mbox{pole}}
\newcommand{\aver}[1]{\langle #1\rangle}

\newcommand{\appa}{\mbox{\ae}}
\newcommand{\CP}{{\em CP } }
\newcommand{\fy}{\varphi}
\newcommand{\hi}{\chi}
\newcommand{\al}{\alpha}
\newcommand{\as}{\alpha_s}
\newcommand{\gf}{\gamma_5}
\newcommand{\gm}{\gamma_\mu}
\newcommand{\gn}{\gamma_\nu}
\newcommand{\be}{\beta}
\newcommand{\ga}{\gamma}
\newcommand{\de}{\delta}
\renewcommand{\Im}{\mbox{Im}\,}
\renewcommand{\Re}{\mbox{Re}\,}
\newcommand{\GeV}{\,\mbox{GeV}}
\newcommand{\MeV}{\,\mbox{MeV}}
\newcommand{\matel}[3]{\langle #1|#2|#3\rangle}
\newcommand{\state}[1]{|#1\rangle}
\newcommand{\ra}{\rightarrow}
\newcommand{\ve}[1]{\vec{\bf #1}}

\newcommand{\rhs}{{\em rhs}}
\newcommand{\pp}{\langle \ve{p}^2 \rangle}

\newcommand{\BR}{\,\mbox{BR}}
\newcommand{\La}{\overline{\Lambda}}

\begin{flushright}
\large{
UND-HEP-94-BIG\hspace*{0.1em}11}\\
November, 1994
\end{flushright}
\vspace{1.2cm}
\begin{center} \LARGE {\bf THE EXPECTED, THE PROMISED AND
THE CONCEIVABLE -- ON CP VIOLATION IN BEAUTY AND CHARM DECAYS
\footnote{Invited talk given at
{\bf HQ 94}, University of Virginia, Charlottesville,
October 1994}}
\end{center}
\begin{center}
{\bf I.I. Bigi}\\

{\it Dept. of Physics,
University of Notre Dame du
Lac, Notre Dame, IN 46556, U.S.A.\\
e-mail address: BIGI@UNDHEP.HEP.ND.EDU}\\
%
\end{center}
\thispagestyle{empty}
\vspace{.4cm}

\centerline{\bf Abstract}
\noindent
The general CP phenomenology for beauty and charm decays is sketched and
the KM expectations of large CP asymmetries in $B$ decays are
reviewed. I describe some observable signatures for the
intervention of New Physics and list benchmarks defining the
`ultimate' measurements in beauty physics. I also stress the need for
dedicated searches for CP asymmetries in $D$ decays; attaining a
sensitivity level of $10^{-3}$ could well reveal New Physics and thus
lead to a new paradigm.
\vspace{1.0cm}

\addtocounter{page}{-1}

\section{-- Introduction}

The three basic messages I want to convey in this talk are
contained in the title:

\noindent $\bullet$ We can confidently {\em expect} large CP asymmetries
to occur in $B$ decays.

\noindent $\bullet$ I feel almost justified to {\em promise} that New
Physics (NP) will reveal itself there.

\noindent $\bullet$ It is quite {\em conceivable} that CP asymmetries will
become observable in charm decays -- in particular if an intervention of
NP will enhance their size above the levels expected within the KM ansatz.

The talk will be organized as follows: the general phenomenology is
sketched in Sect. 2 and the KM expectations for $B$ decays are stated in
Sect. 3; signatures for NP are listed in Sect. 4; Sect. 5 contains the
`HERA-B menu' while the `ultimate' measurements are defined in Sect. 6;
after describing the most promising ways to search for CP violation in
charm decays in Sect. 7, I give an outlook in Sect. 8.

\section{-- General Phenomenology of CP Violation}

There are five different classes of CP asymmetries that can emerge in
meson decays, and they fall into two groups.

The first group involves comparing partial rates. The decay
rate for two CP conjugate channels as a function of proper time $t$ can
be written as
$$\Gamma (B[D]\ra f;t)=e^{-\Gamma _{B[D]}t}G_f(t)\, , \; \; \;
\Gamma (\bar B[\bar D]\ra \bar f;t)=
e^{-\Gamma _{B[D]}t}\bar G_{\bar f}(t)\eqno(2.1)$$
If $\bar G_{\bar f}(t)/G_{f}\neq 1$
is observed, CP violation has been discovered. Three classes can be
distinguished:

\noindent (a):
$$\frac{\bar G_{\bar f}(t)}{G_f(t)}\neq 1 \, , \; \; \;
\frac{d}{dt}\left( \frac{\bar G_{\bar f}(t)}{G_f(t)}\right) \neq 0
\eqno(2.2)$$
This represents `CP violation involving $B^0-\bar B^0$[$D^0-\bar D^0$]
oscillations and it can manifest itself in non-leptonic decays: e.g.,
$B_d\ra \psi K_S$ or $D^0\ra K^+K^-,\, \pi ^+\pi ^-$
\cite{CBS}.

\noindent (b):
$$\frac{\bar G_{\bar f}(t)}{G_f(t)}\neq 1 \, , \; \; \;
\frac{d}{dt}\left( \frac{\bar G_{\bar f}(t)}{G_f(t)}\right)
\equiv 0,\;
\frac{d}{dt}G_f(t)\neq 0\eqno(2.3)$$
It constitutes `CP violation in $B^0-\bar B^0$[$D^0-\bar D^0$]
oscillations' which can most clearly be studied in semileptonic decays,
in particular to `wrong-sign' leptons:
$B^0=(b\bar q)\ra l^+X$ or $D^0\ra l^-X$.

\noindent (c):
$$\frac{\bar G_{\bar f}(t)}{G_f(t)}\neq 1 \, , \; \; \;
\frac{d}{dt}\left( \frac{\bar G_{\bar f}(t)}{G_f(t)}\right)
\equiv 0,\;
\frac{d}{dt}G_f(t)\equiv 0\eqno(2.4)$$
This case is called `direct' CP violation and it can also arise in
charged meson and baryon decays: e.g.,
$B\ra K\pi ,\, D^{neut}K$ or $D\ra K_S\pi$.

The second group involves observables other than a ratio
of partial widths.

\noindent (d): CP violation can reveal itself in final-state distributions:
there can be telling asymmetries in
the Dalitz plot, or T odd correlations can arise.

\noindent (e): CP violation can be established also through the observation of
a
special transition rather than an asymmetry. Consider
$$e^+e^- \ra B^0\bar B^0[D^0\bar D^0]
|_{J^{CP}=1^{--}}\ra f_1f_2 \eqno(2.5)$$
with $f_1$ and $f_2$ being CP eigenstates of the
{\em same} CP parity. Reaction
(2.5) can proceed only through CP violation since the initial and
final state CP parities differ.

The prospects for observing such phenomena vary considerably
from case to case: class $(e)$ will presumably remain academic since its
rate depends on the product of two small branching ratios like
for $f_{1,2}=\psi K_S\ra (l^+l^-)K_S$ or $D^+D^-\ra (K^-\pi ^+\pi ^+)
(K^+\pi ^-\pi ^-)$. Class $(d)$ represents a very wide and promising
field; some interesting theoretical studies have been made \cite{KRAMER},
but it is still
too early to draw firm conclusions, and I will not pursue it any further here.
As far as class $(b)$ is concerned, existing predictions are not very precise,
but it is very hard to see how CP asymmetries in semileptonic $B^0$ decays
could exceed 0.1 \%.

In the following I will focus on classes $(a)$ and $(c)$, which have
complementary advantages and drawbacks, as sketched in
Table \ref{TABLEI}.

\begin{table}
\centering
\caption{}
\begin{tabular} {|l|l|l|}
\hline
& &\\
Class &Advantages&Drawbacks \\
& & \\
\hline
\hline
&  & \\
CP violation &
$\bullet$ reliable theoretical predictions
&$\bullet$ flavour tagging required\\
involving
&$\bullet$ large effects &
$\bullet$ asymmetric $B$ factory\\
oscillations
&$\bullet$ striking experimental signature& needed  \\
& $\bullet$ access to all three angles of the & \\
& KM triangle & \\
& & \\
\hline
& & \\
direct CP & $\bullet$ self-tagging &
$\bullet$ less clean theoretical
\\
violation
&$\bullet$ can be done at a symmetric & interpretation\\
& $B$ factory &$\bullet$ smaller effects \\
&  & $\bullet$ less striking experimental\\
& & signature \\
&  & $\bullet$ access to only one angle\\
& & in the KM triangle \\
& & \\
\hline
\end{tabular}
\label{TABLEI}
\end{table}

\section{-- KM Expectations in B Decays}

\subsection{\underline {Generalities}}
Within the KM ansatz the CP asymmetries are described in terms of
relative phases of various KM parameters.
Weak universality imposes unitarity constraints on them that are
expressed by triangle relations\cite{JARLSKOG}:
$$ 0=V^*(ud)V(ub)+V^*(cd)V(cb)+V^*(td)V(tb)\simeq V(ub)-
\lambda V(cb) + V^*(td), \, \lambda \simeq \sin \theta _C
\eqno(3.1)$$
Since the three sides of this triangle are all
$\sim {\cal O}(\lambda ^3)$, its angles are naturally large,
see Fig.~\ref{F1}.
\begin{figure}
\begin{center}
\psfig{figure=7207fg1.eps,height=5cm}}
\vspace*{-1cm}
\end{center}
\caption[]{The KM triangle relevant for $B$ transitions}
\label{F1}
\end{figure}

This is in
marked contrast to the situation in charm decays where the relevant
triangle is given by
$$ 0=V^*(ud)V(cd)+V^*(us)V(cs)+V^*(ub)V(cb)\simeq
{\cal O}(\lambda )+{\cal O}(\lambda ) + {\cal O}(\lambda ^5)
\eqno(3.2)$$
which represents a very `squashed' triangle with one angle necessarily
tiny and the other two accessible only through highly suppressed
transitions with an amplitude of relative strength
${\cal O}(\lambda ^4)\sim 3\cdot 10^{-3}$.

The foremost goal is of course to find CP violation. Yet one can define
a more specific and detailed program of inquiry, namely to
first determine and then probe the KM triangle with utmost
sensitivity. For that purpose one aims at extracting the values of
the KM parameters $|V(cb)|$, $|V(ub)|$ and $|V(td)|$ from various
sets of data\cite{NAGOYA}. For these quantities determine the
lengths of the three
sides of the relevant KM triangle and thus allow to infer the
values of the angles. Next one undertakes to measure the angles
$\phi _1$ and $\phi _2$ (also known as $\beta$
and $\al$). If their  measured values differed significantly from their
infered values, one would have established the intervention of
New Physics (NP).
There is a clear prescription of how to measure these angles through
CP asymmetries. Consider for simplicity only final states $f$ that are CP
self-conjugate; one then obtains
$$\Gamma (B^0[\bar B^0]\ra f;t)=K_f[\bar K_f]e^{-\Gamma _Bt}
\left( 1-[+]Im\frac{q}{p}\bar \rho _f\cdot \sin \Delta m_Bt\right) ,\;
\bar \rho _f=\frac{T(\bar B^0\ra f)}{T(B^0\ra f)}\eqno(3.3)$$
For $f=\psi K_S$ one finds that the asymmetry parameter can be
expressed -- to a high degree of accuracy --
as a combination of KM parameters only:
$(q/p)\cdot (T(\bar B^0\ra \psi K_S)/T(B^0\ra \psi K_S))=
(V^*(td)/V(td))\cdot (V^*(cb)V(cs)/V(cb)V^*(cs))$;
therefore
$$\Gamma (B_d[\bar B_d]\ra \psi K_S;t)=K_1e^{-\Gamma _Bt}
\left( 1-[+]\sin 2\phi _1 \sin \Delta m_Bt\right) \eqno(3.5)$$
Similarly
$$\Gamma (B_d[\bar B_d]\ra \pi ^+\pi ^-;t)=
K_2[\bar K_2]e^{-\Gamma _Bt}
\left( 1-[+]"\sin 2\phi _2" \sin \Delta m_Bt\right) \eqno(3.6)$$
As far as $B_d\ra \pi ^+\pi ^-$ is concerned there arise two
complications. On the one hand the quantity $"\sin 2\phi _2"$ measured in
$B_d\ra \pi ^+\pi ^-$ is not identical to the genuine KM parameter
$\sin 2\phi _2$. The difference between the two can be ascribed to
Cabibbo suppressed Penguin transitions $b\ra d$ or to rescattering
$B_d\ra "D\bar D"\ra \pi ^+\pi ^-$. While I do not expect this
difference to be big, it exists and provides a limiting factor in the
theoretical interpretation. This added complexity will typically (though
not necessarily) also lead to direct CP violation in $B_d\ra \pi ^+\pi ^-$
decays as allowed in eq. (3.6) through $K_2\neq \bar K_2$. Measuring
$BR(B_d\ra \pi ^0\pi ^0)$ and $BR(B^-\ra \pi ^-\pi ^0)$ will however
enable us to extract $\sin 2\phi _2$ reliably from the CP asymmetry in
$B_d\ra \pi ^+\pi ^-$\cite{PENGUINS}.
In addition to this theoretical complication there exists an experimental
one as well: it is obviously important to have particle identification
that can reliably distinguish $B\ra K\pi$ and $B\ra \pi \pi$ modes.

There is also some good news which I will only state: there exists a host
of additional exclusive channels that can be employed here:
$B_d\ra \psi K_L$, $\psi ' K_S$, $\psi 'K_L$, $\psi (K_S\pi ^0)_{K^*}$,
$D^{(*)}\bar D^{(*)}$ etc. for $\sin 2\phi _1$ and $B_d\ra \pi \rho , \,
\pi \omega , \, \pi a_1$ etc. for $\sin 2\phi _2$.

\subsection{-- Determining the KM Triangle}
The baseline of the triangle can conveniently be normalized to unity
without affecting the angles. The other two sides are then given by
$V(ub)/V(cb)$ and $V^*(td)/\lambda V(cb)$; those ratios are more directly
observable than $V(ub)$ and $V(td)$ themselves.

Our present information on the normalized KM triangle is as follows:

\noindent $\bullet$ Present data on charmless semileptonic $B$ decays
{\em suggest} $|V(ub)/V(cb)|\simeq 0.08 \pm 0.03$. I am somewhat skeptical that
the stated error properly reflects the
present experimental and theoretical
uncertainties;
yet I am confident that the situation will be clarified in the next
few years due to considerably more sensitive data and a judicious application
of heavy quark expansions\cite{NAGOYA}.

\noindent $\bullet$ The quantity $\sin 2\phi _1$ controling the CP asymmetry in
$B_d \ra \psi K_S$ can in principle be extracted from the observed value of
$\epsilon _K/\Delta m(B_d)$ with little sensitivity to the
(heavy) top quark mass. Yet this procedure at present suffers from grave
theoretical uncertainties in the size of the hadronic matrix element:
$$\frac{\epsilon _K}{\Delta m(B_d)}\propto \sin 2\phi _1 \simeq
0.42\cdot UNC \eqno(3.7a)$$
$$UNC\simeq \left( \frac{0.04}{|V(cb)|}\right) ^2
\left( \frac{0.72}{x_d}\right)
\left( \frac{\eta _{QCD}^{(B)}}{0.55}\right)
\left( \frac{0.62}{\eta _{QCD}^{(K)}}\right)
\left( \frac{2B_B}{3B_K}\right)
\left( \frac{f_B}{160\, \MeV}\right) ^2
\eqno(3.7b)$$
Nevertheless two things should be noted here:

\noindent -- $\sin 2\phi _1$ is {\em large} for reasonable values of
$f_B$; for $f_B=220$ MeV one actually obtains $\sin 2\phi _1 \simeq 0.8$!

\noindent -- Once $\sin 2\phi _1$ has been measured, one can infer the
value required for $f_B$ from eqs.(3.7) with good accuracy.

The present information on the KM triangle is summarized in
Fig.~\ref{F4}:
the top of the triangle has to lie in the shaded area.
\begin{figure}
\begin{center}
\psfig{figure=7207fg4.eps,height=5cm}}
\vspace*{-1cm}
\end{center}
\caption[]{Shape of the KM triangle inferred from present
phenomenology}
\label{F4}
\end{figure}

I anticipate
that over the next few years $|V(ub)/V(cb)|$ will be extracted with
a realistic error of 10\% or less. It is hoped that the
top quark mass will be known to within $\pm 10$ GeV. The resulting
KM landscape is illustrated in
Fig.~\ref{F5}.
\begin{figure}
\begin{center}
\psfig{figure=7207fg5.eps,height=5cm}}
\vspace*{-1cm}
\end{center}
\caption[]{Shape of the KM triangle after future measurements of
$V(ub)/V(cb)$ and $m_t$}
\label{F5}
\end{figure}
The allowed area for the top of the
triangle is now greatly reduced, and consists of two disjoint
subdomains; one requires $f_B\simeq 210\div 240$ MeV and the other
$f_B\simeq 140\div 170$ MeV.

Knowing $|V(td)/V(cb)|$ and thus the third side would provide another
powerful constraint; yet that information will not come easily.
There are three avenues towards this goal: (i) It has
been suggested \cite{ALI} to
extract it from the observed ratio $R_{\gamma}\equiv
BR(B\ra \gamma \rho /\omega)/BR(B\ra \gamma K^*)$. This is based on
the assumption that both radiative transitions are driven mainly by a
Penguin operator reflecting short distance dynamics; in that case
one would have $R_{\gamma}=|V(td)/V(ts)|^2\times SU(3)$ breaking.
Unfortunately it has not been established that in particular the
mode $B\ra \gamma \rho /\omega$ is Penguin dominated. One can
actually advance various arguments why long distance dynamics make quite
significant contributions that do not depend on $V(td)$. I am skeptical
that such contributions can reliably be computed from first principles in the
near future; on the other hand one can gauge their weight once
$BR(D\ra \gamma K^*,\, \gamma \rho /\omega)$ and $BR(B\ra \gamma D^*)$ have
been measured
since these modes cannot be driven by Penguin operators\cite{MANNEL}.
(ii) Determining the $B_s-\bar B_s$ oscillation rate
would allow a fairly reliable extraction:
$$\frac{\Delta m(B_d)}{\Delta m(B_s)}\simeq
\frac{Bf_B^2|_{B_d}}{Bf_B^2|_{B_s}}\cdot \frac{|V(td)|^2}{|V(ts)|^2}
\eqno(3.8)$$
where it is understood that the {\em ratio} of the $B_d$ and $B_s$
matrix elements can be calculated more reliably than the matrix elements
themselves. (iii) My own favourite is to employ
a measurement of $BR(K^+\ra \pi ^+\nu \bar \nu)$ once that becomes
available. For the width of that rare transition can reliably
be expressed as a function of $m_{top}$ and $|V(td)|$ with the
remaining uncertainty mainly due to the size of $m_{charm}$
\cite{BURAS,GABBIANI1}.

\section{-- Signatures for New Physics}

The best way in which searches for NP are to be conducted will
depend on how much is known at the time about which KM
parameter. I will describe here two typical scenarios to illustrate
the basic features on which any search would be based.
\subsection{\underline {Typical Search Scenarios}}
(1) With $|V(ub)/V(cb)|$ and $|V(td)/V(cb)|$ known the KM
triangle has been determined. Actually it would already have been
overconstrained without data on CP violation in $B$ decays:
for with eqs.(3.7) one can infer the necessary size of $f_B$
from the resulting angle $\phi _1$. With it and
the measured top mass one computes $\Delta m(B_d)$ and confronts it
with the experimental value. Measuring the CP asymmetry in
$B_d\ra \psi K_S$ then yields a second sensitive constraint: if it
is found to
differ from $\sin 2\phi _1$ as infered from the triangle, one has
established the intervention of NP.

(2) If $V(td)/V(cb)$ were remain to be largely undetermined, one had to
use the measured CP asymmetry in $B_d\ra \psi K_S$ to fix $\sin 2\phi _1$
and thus the triangle. As before it would then actually have been
overdetermined: inferring the size of $f_B$ from $\sin 2\phi _1$ one
can compute $\Delta m(B_d)$ and compare it with the data. Measuring
$\sin 2\phi _2$ to obtain a second constraint would then become even
more mandatory.
\subsection{\underline {The $\phi _3$ (or $\gamma$) Saga}}
The third angle $\phi _3$ can be measured via CP violation involving
$B^0-\bar B^0$ oscillations or via direct CP violation. The first method
searches for a CP asymmetry in, e.g., $B_s\ra K_S \rho ^0$ decays. However
it does not strike me as particularly promising. For by then the other two
angles $\phi _1$ and $\phi _2$ will have been determined
with a higher accuracy than can
realistically be expected for $\phi _3$ measured in this way; $\phi _3$ will
then be known as $\pi -\phi _1 -\phi _2$ -- unless there is NP lurking below
the
surface! Yet then it makes more sense to analyze a channel with (a) a higher
branching ratio, (b) a more striking experimental signature and (c) a cleaner
theoretical interpretation. The mode $B_s\ra \psi \phi$
(or $B_s\ra \psi \eta$) fits this bill\cite{BS}.
For the KM ansatz predicts a small
asymmetry here
$$Im \frac{q}{p}\bar \rho (B_s\ra \psi \phi)|_{KM}\sim 2\% \eqno(4.1)$$
as is easily understood: for on the leading KM level
only quarks of the second and third family contribute in the transitions
$B_s\ra \psi \phi$ and $B_s\ra \bar B_s\ra \psi \phi$; a CP asymmetry then
has to be Cabibbo suppressed. New Physics on the other hand could quite
naturally produce an asymmetry well in excess of 10\%! A note of caution: if
$B_s$ mesons oscillated too rapidly, the asymmetry in $B_s\ra \psi \phi$
would get washed out -- yet this would also happen in
$B_s\ra K_S\rho ^0$.

The angle $\phi _3$ can be measured also via a direct CP asymmetry in
$B^{\pm}\ra D_{neut}K^{\pm}$ modes\cite{PAIS}.
This represents a sounder
approach, in particular since it can be undertaken already at a
{\em symmetric} $B$ factory. Furthermore it is at least
intellectually `cute'. For it makes use of the fact that
for neutral mesons flavour eigenstates and mass eigenstates are
in general distinct (although they can be expressed as linear combinations
of each other) with the former
defined by the
flavour-specific decays $D^0\ra K^-\pi ^+$ and $\bar D^0\ra K^+\pi ^-$
\footnote{Due to doubly Cabibbo suppressed decays these channels do
not provide a perfect filter, yet one of sufficient quality.} and the
latter by decays into CP eigenstates: $D_1\ra K_S\pi ^0$,
$D_2\ra K_L\pi ^0, \, K^+K^-$. A comparison of
$B^-\ra D_1K^-[D_2K^-]$ with $B^+\ra D_1K^+[D_2K^+]$ can then reveal
a CP asymmetry for these channels reflecting the coherent interplay
of $B^{\pm}\ra D^0K^{\pm}$ and $B^{\pm}\ra \bar D^0K^{\pm}$
transitions. It should be noted as a point of general interest
that the quantum mechanical realization
of particle-antiparticle identity for $B^0$, $D^0$ and $K^0$ mesons
plays an essential role in CP asymmetries of $B$ decays: (i) Applied
to $B^0$ states it constitutes a pre-requisite for $B^0-\bar B^0$
oscillations. (ii) It leads to distinct states $K_S$ and $K_L$ to
provide the second pre-requisite for the observability of a CP
asymmetry in $B_d\ra \psi K_S$. (iii) Applied to $D^0$ mesons it
is essential for the emergence of direct CP violation in
$B_d\ra DK$ decays, as dicussed above.

The latter is proportional to $\sin \phi _3$ \footnote{It is also
amusing to note that $\sin \phi _3 =1$, $\sin 2\phi _3 =0$ for
$\phi _3 = 90^o$ as implied by the `Stech' matrix.}; it also depends
on the phaseshift
$\Delta \phi _{str}$. Yet it is a gratifying feature
\cite{GRONAU} of these
reactions that they allow the experimental isolation of $\sin \phi _3$
from the hadronic matrix elements.
Due to CPT invariance one has four independant
amplitudes, namely for $B^-\ra D^0K^-$, $B^-\ra \bar D^0K^-$,
$B^-\ra D_{1,2}K^-$ and $B^+\ra D_{1,2}K^+$. Measuring them allows one
to extract $\sin \phi _3$, $\Delta \phi _{str}$, $|T(B^-\ra D^0K^-)|$ and
$|T(B^-\ra \bar D^0K^-)|$ (up to a discrete ambiguity)!
\subsection{\underline {New Physics Scenarios}}
One can easily give explicit dynamical scenarios of NP that would have an
observable impact on the CP phenomenology in $B$ decays.

As usual, the Minimal Supersymmetric Standard Model (MSSM) plays the role of
the standard extension of the Standard Model. Since it does not introduce any
appreciable new phase, there is still {\em no} sizeable CP asymmetry in
$B_s\ra \psi \phi$. Yet MSSM can have an indirect impact
\cite{GABBIANI2}. For it can
provide a significant contribution to $\Delta m(B_d)$ and
$\Delta m(B_s)$, though not to their ratio. With
$\Delta m(B_d)/\Delta m(B_s)$ and thus $|V(td)/V(ts)|$ measured in addition to
$|V(ub)/V(cb)|$ one would observe the following in the MSSM scenario: the
angles
$\phi _1$ and $\phi _2$ as infered from the KM trigonometry would agree with
the
observed CP asymmetries in $B_d\ra \psi K_S$ and $B_d\ra \pi ^+\pi ^-$; yet
with the value infered for $f_B$ from $\sin 2\phi _1$ one would fail to
reproduce
$\Delta m(B_d)$! The discrepancy would be due to SUSY contributions.

A {\em non-minimal} implementation of SUSY on the other hand would open the
floodgates for additional CP phases\cite{GABBIANI2}
-- as would horizontal interactions etc.
etc. In general NP is most likely to make its presence felt first in the
highly forbidden $\Delta B=2$ transitions driving $B^0-\bar B^0$ oscillations.
In that case it would affect all $B_d$ decays in a uniform manner; likewise for
$B_s$ decays. It would thus represent a dynamical realization of the
`Superweak Scenario'.

\section{-- The HERA-B Menu}

Contrary to widely held beliefs good food can be found at DESY.
I am actually referring to food for the mind that is being offered
on the HERA-B menu. It consists of three main courses:

\noindent (i) $B_s-\bar B_s$ oscillations,

\noindent (ii) $B_d\ra \psi K_S$ vs. $\bar B_d\ra \psi K_S$,

\noindent (iii) $B_s\ra \psi \phi$ vs. $\bar B_s\ra \psi \phi$,

\noindent and one side dish:

\noindent (iv) $\tau (B_s \ra l \nu D_s^{(*)})$ vs.
$\tau (B_s\ra \psi \phi)$.

Observing a positive signal in all three cases (i) - (iii) would
represent a `Beyond your wildest dreams' scenario: as discussed before,
the results from (i) would yield the final element for the KM triangle
together with one constraint from $\Delta m(B_d)$; course (ii) would
provide a second constraint. Taken together (i) and (ii) represent
probes for NP in $\Delta m(B^0)$ and Im$\Delta m(B_d)$. Course (iii)
constitutes a clean and direct probe for NP in Im$\Delta m(B_s)$.
Concerning the side dish (iv): $B_s-\bar B_s$ oscillations generate
two states differing also in their lifetimes: $B_s^{(short)}$ and
$B_s^{(long)}$. The $B_s^{(short)}$ lifetime reveals itself, to a good
approximation, in $B_s\ra \psi \phi$, whereas the semileptonic channels
$B_s\ra l \nu D_s^{(*)}$ reflect the lifetime averaged over
$B_s^{(short)}$ and $B_s^{(long)}$. A state of the art calculation yields
\cite{GAMMABS}
$$\frac{\tau (B_s\ra l\nu D_s^{(*)})-\tau (B_s\ra \psi \phi )}
{\tau (B_s\ra l\nu D_s^{(*)})}\simeq 0.1 \cdot
\left( \frac{f_{B_s}}{200\, \MeV}\right) ^2 \eqno(5.1)$$
Such a small difference in lifetimes would escape detection. Yet it is
conceivable that the real difference is considerably larger since
the computation yielding eq.(5.1) is not gold-plated. Searching for the
lifetime difference in $B_s^{(short)}$ vs. $B_s^{(long)}$ thus represents
a probe not for NP, but of our computational control over hadronization
effects.

\section{-- The Ultimate Challenge in $B$ Physics}

There is good reason for the hope that the asymmetric $B$ factories at
KEK and SLAC will establish CP violation in $B$ decays. Nevertheless
it is unlikely that they will provide also the ultimate measurements. Those
are defined by the following considerations:

(i) Some of the theoretical predictions -- like the asymmetry parameter in
$B_d\ra \psi K_S$ being $\sin 2\phi _1$ -- enjoy a high parametric
accuracy. Furthermore there exists the expectation that an increasingly
detailed database coupled with refined theoretical tools will allow us
to translate this parametric accuracy into a numerical one of a few
per cent\cite{NAGOYA}.
This opens up the possibility to search for NP contributing
as little as 10\% in amplitude.

(ii) It is quite possible that no CP violation were found in $B$ decays, with
an
upper bound of, say, a very few per cent. In that case we would have
established that $K_L\ra \pi \pi $ transitions are predominantly driven by a
source (or sources) other than the KM mechanism, i.e NP!

(iii) Even finding the CP phenomenology in $B$ decays to be fully
consistent with the KM framework could provide us with seeds of more
profound knowledge: analyses of SUSY GUT scenarios lead to an apparently
rather limited set of allowed KM matrices. Prominent
examples \cite{RRR,RABY} are listed in
Table \ref{TABLEII} for $m_{top}=180\, \GeV$
in a way that is convenient for my discussion,
where I have assumed $[B_sf^2(B_s)]/[B_df^2(B_d)]=1.1$
to translate $|V(ts)/V(td)|^2$ into $x_s/x_d$.
\begin{table}
\centering
\caption{Possible scenarios for the KM matrix for
$m_{top}=180$ GeV.}
\begin{tabular} {|l|l|l|l|l|l|l|l|l|}
\hline
&A &B &C &D &E &I &II &III \\
\hline
\hline
$|V(ub)/V(cb)|$ & 0.06 & 0.062 & 0.068 & 0.059
&0.089 &0.046 &0.059 &0.071 \\
\hline
$x_s/x_d$ & 25 & 25 & 22 & 26 & 35 &  &  &  \\
\hline
$\sin 2\phi _1$ & 0.52 & 0.54 & 0.58 & 0.51 & 0.71
&0.39  &0.49  &0.59  \\
\hline
$\sin 2\phi _2$ & $-$0.15 & $-$0.11 & 0.30 & $-$0.40 &
$-$0.62 &$-$0.32 &$-$0.46 &$-$0.14 \\
\hline
$\sin 2\phi _3$ & 0.65 & 0.71 & 0.30 & 0.81 & 0.99
&0.66 &0.84 &0.70 \\
\hline
$\sin \phi _3$ & 0.94 & 0.94 & 0.99 & 0.89 & 0.75
&0.93 &0.88 &0.93 \\
\hline
\end{tabular}
\label{TABLEII}
\end{table}
The symbols
$A-E$ and $I-III$ refer to different classes of mass matrices
analysed in ref. \cite{RRR} and
\cite{RABY}.  The details are not important here
\footnote{It takes, of course, the trained eye of a theorist
to discern the simple pattern underlying these values for the
KM parameters.},
and I anticipate considerable theoretical evolution to take place
over the next few years; but I want to use these numbers to
illustrate important benchmarks for the ultimate measurements:

\noindent $\bullet$ $B_s - \bar B_s$ oscillations might be
extremely rapid!

\noindent $\bullet$ The business at hand remains unfinished in an
essential way until also $\sin 2\phi _2$ has been measured with good accuracy.

\noindent $\bullet$ To differentiate completely among these
scenarios  requires the experimental
uncertainties to lie below a few per cent.

Obviously only dedicated experiments performed at the LHC have the
statistical muscle to achieve such goals. The question is:  can
they develop the systematic brain that is up to this task?

\section{-- CP Violation in Charm Decays}

It has been stated quite often that without NP $D^0-\bar D^0$
oscillations proceed very slowly and CP asymmetries in $D$ decays
are small. Yet: How small is small? This has
to be addressed on a case-by-case basis.
\subsection{\underline {CP Violation Involving
$D^0-\bar D^0$ Oscillations}}
The rate for $D^0$ decays into CP eigenstates like $K^+K^-$
or $\pi ^+\pi ^-$ as a function of proper time $t$ is given by
$$\Gamma (D^0\ra f;t)\propto e^{-\Gamma _Dt}
\left( 1+\sin \Delta m_Dt \, Im\frac{q}{p}
\bar \rho (D\ra f)\right) \simeq$$
$$\simeq e^{-\Gamma _Dt}
\left( 1+\frac{\Delta m_D}{\Gamma _D}\frac{t}{\tau _D}
Im\frac{q}{p}\bar \rho (D\ra f)\right) \eqno(7.1)$$
In the Standard Model one expects $\Delta m_D/\Gamma _D\sim 0.01$
and $Im\bar \rho (D\ra f)\sim {\cal O}(\lambda ^4)\leq 0.01$;
i.e. such a CP asymmetry will not exceed the $10^{-4}$ level, and it
is so tiny since it represents the product of two small effects, namely
$D^0-\bar D^0$ oscillations and CP violation.
\subsection{\underline {Direct CP Violation}}
Direct CP violation can become observable only if two different amplitudes
contribute coherently to a certain decay. This can happen for once
Cabibbo suppressed modes, but neither for Cabibbo allowed nor twice
forbidden channels. Rough estimates tell us that direct CP asymmetries
can be as `large' as ${\cal O}(\lambda ^4)\sim {\cal O}(10^{-3})$ in
$D\ra [S=0]$ and $D_s\ra [S=+1]$ channels. It will be possible to
refine these predictions in the future through `theoretical engineering'
\cite{BUCCELLA},
i.e. one matches theoretical predictions for two-body modes against
a host of well-measured branching ratios to extract the size of
transition amplitudes (including absorption) and strong phase shifts.
Present data limit direct CP asymmetries to roughly the 10\% level,
as shown in Table \ref{TABLEIII}.

\begin{table}
\centering
\caption{Present bounds on CP asymmetries in $D$ decays}
\begin{tabular}{|l|l|l|}
\hline
 Decay mode & Measured asymmetry & 90\% C.L. limit \\
\hline
\hline
$D^0\rightarrow K^+K^-$ & $0.024\pm 0.084$ \cite{E687}&
$-11\%<A_{CP}<16\%$ \\
 &$0.071 \pm 0.065$ \cite {CLEOII}  &$-3.6\% <A_{CP}<17.8\%$ \\
\hline
$D^+\rightarrow K^-K^+\pi ^+$ & $-0.031\pm 0.068$ \cite{E687}
& $-14\%<A_{CP}<8.1\%$ \\
\hline
$D^+\rightarrow \bar K^{*0}K^+$ & $-0.12\pm 0.13$ \cite{E687}&
$-33\%<A_{CP}<9.4\%$ \\
\hline
$D^+\rightarrow \phi \pi ^+$ & $0.066\pm 0.086$ \cite{E687}&
$-7.5\%<A_{CP}<21\%$ \\
\hline
$D^0\rightarrow K_S\phi$ &$-0.005\pm 0.067$ \cite{CLEOII}
&$-11.5\%<A_{CP}<10.5\%$ \\
\hline
$D^0\rightarrow K_S\pi ^0$ &$-0.011\pm 0.030$ \cite{CLEOII}
&$-6\%<A_{CP}<3.8\%$ \\
\hline
\end{tabular}
\label{TABLEIII}
\end{table}
\vspace{0.5cm}

\subsection{\underline {Possible Impact of NP}}
The intervention of NP could have a two-fold impact on CP violation
involving $D^0-\bar D^0$ oscillations: it could quite conceivably
increase $\Delta m_D/\Gamma _D$ to around $0.1$, its present upper
bound; secondly, it could generate $Im(D^0\ra K^+K^-, \, \pi ^+\pi ^-)
\sim 0.1$. Taken together, it would produce a CP asymmetry
$\sim {\cal O}(1\%)$ in reaction (7.1).

A new element overlooked so far emerges for direct CP asymmetries in
$D\ra K_S+\pi 's$ transitions\cite{YAMAMOTO}. I will describe it here
specifically
for $D^{\pm}\ra K_S\pi ^{\pm}$ decays. There are two different amplitudes
contributing to this mode, namely the Cabibbo allowed channel
$D^+\ra \bar K^0\pi ^+$ and the doubly Cabibbo suppressed one (DCSD)
$D^+\ra K^0\pi ^+$:
$$\Gamma (D^+\ra K_S\pi ^+)\simeq \frac{1}{2}\Gamma (D^+\ra \bar K^0\pi ^+)
\left( 1-2Re\frac{T(D^+\ra K^0\pi ^+)}{T(D^+\ra \bar K^0\pi ^+)}
\right) \eqno(7.2)$$
Four consequences arise from the coherent contribution of the DCSD amplitude:

\noindent $\bullet$ $\Gamma (D^+\ra K_S\pi ^+)
\neq \Gamma (D^+\ra K_L\pi ^+)$ even
ignoring CP violation in the $K^0-\bar K^0$ system.

\noindent $\bullet$ The difference is linear in the DCSD amplitude and
thus of order $2\tan ^2\theta _C=0.1$.

\noindent $\bullet$ With the isospin structure differing for
$D^+\ra \bar K^0\pi ^+$ and $D^+\ra K^0\pi ^+$ one can expect
different strong phase shifts to occur in the two amplitudes. Those
can actually be determined experimentally. Thus a CP asymmetry
can emerge:
$$\Gamma (D^+\ra K_S\pi ^+)
\neq \Gamma (D^-\ra K_S\pi ^-)\eqno(7.3)$$
\noindent $\bullet$ An asymmetry of equal size, but opposite sign
occurs for the final state where $K_L$ replaces $K_S$:
$$\Gamma (D^+\ra K_L\pi ^+)-\Gamma (D^-\ra K_L\pi ^-)=
-[\Gamma (D^+\ra K_S\pi ^+)-\Gamma (D^-\ra K_S\pi ^-)]\eqno(7.4)$$
Such an asymmetry arises already within the KM ansatz. Yet it
is tiny, namely $\leq 10^{-4}$, again reflecting the product of two
small quantities $\sim \lambda ^2\cdot \lambda ^4$. Yet if NP
contributes a mere 10\% to the DCSD amplitude, one would have an
asymmetry of around 1\%!

\section{-- Outlook}

It is quite evident that the insights to be gained from a comprehensive
and dedicated study of CP violation

\noindent $\bullet$ are of fundamental importance;

\noindent $\bullet$ cannot be obtained any other way, and

\noindent $\bullet$ cannot become obsolete.

The phenomenon of CP violation can be put also into a wider
context. There are two central mysteries in the Standard
model. One concerns the origin of masses: while the
generation of mass can be implemented by the Higgs mechanism in
a gauge invariant way, there is no direct experimental evidence for
it; furthermore it is also quite unsatisfactory from a theoretical
point of view. The second central mystery -- not unrelated to the
first one -- concerns the family replication, the pattern of the
fermion masses and the origin of CP non-invariance. There we are
even more at a loss for a real understanding; we can only state
that since there are three families, CP violation can be
implemented via the KM mechanism. In short our answer to the question of
why there are families and why there is CP violation is --
`why not?'

There are certainly enough mysteries to ponder. It would be wonderful
if they could be solved by pure thinking -- and preferably all
of them in one fell swoop! Indeed, it would be miraculous. For the
history of our discipline teaches us that progress occurs through
a succession of paradigms with each new one encompassing the previous
one and the shifts most of the time being caused by unexpected
empirical input.

It is actually the motivation for the LHC to gain new insights into
the problem of mass generation by directly probing physics at the
1 TeV scale. Likewise a thorough analysis of CP violation in heavy
flavour decays will provide new perspectives onto the family
problem in general and CP violation in particular. I for myself have
little doubt that these studies will lead to a new paradigm -- in
particular if we commit ourselves to a truly comprehensive
analysis, i.e. one that includes detailed studies of the charm
system (and the $\tau$ lepton and top quark for that matter).

\vspace*{0.5cm}

{\bf Acknowledgements} \hspace{.4cm}
This work was supported by the National Science Foundation under
grant number PHY 92-13313. I enjoyed the format
and the efficient running of the meeting created by the organizers.
It is also always a pleasure to visit the house that `Old Tom'
built!

\vspace*{3cm}

\begin{thebibliography}{17}

\bibitem{CBS}
A. Carter, A. Sanda, Phys. Rev. {\bf D23} (1981) 1567;
I.I. Bigi, A. Sanda, Nucl. Phys. {\bf B193} (1981) 85;
I.I. Bigi, V. Khoze, N.G. Uraltsev, A. Sanda, in:
"CP Violation", C. Jarlskog (ed.), World Scientific, Singapore,
1989, p.175.

\bibitem{KRAMER}
see, e.g.:
G. Kramer, W. Palmer, H. Simma, Nucl. Phys. {\bf B428} (1994) 77.

\bibitem{JARLSKOG}
For a general introduction, see the article by C. Jarlskog, in:
"CP Violation", C. Jarlskog (ed.), World Scientific, Singapore,
1989.

\bibitem{NAGOYA}
I.I. Bigi, Invited Talk given at the International Workshop on B
Physics, Nagoya, Japan, Oct. 26-28, 1994, preprint UND-HEP-94BIG12.

\bibitem{PENGUINS}
M. Gronau, Phys. Lett. {\bf B265} (1991) 389.

\bibitem{ALI}
A. Ali et al., preprint CERN-TH.7118/93, to appear
in Z. Phys. C.

\bibitem{MANNEL}
I.I. Bigi, T. Mannel, in preparation.

\bibitem{BURAS}
A. Buras et al., Phys. Rev. {\bf D50} (1994) 3433.

\bibitem{GABBIANI1}
I.I. Bigi, F. Gabbiani, Nucl.Phys. {\bf B367} (1991) 3.

\bibitem{BS}
I. I. Bigi, A. I. Sanda, Nucl. Phys. {\bf B193}
(1981) 85; Phys. Rev. {\bf D29} (1984) 1393;
Nucl. Phys. {\bf B281} (1987) 41.

\bibitem{PAIS}
I. I. Bigi, A. I. Sanda, Phys. Lett. {\bf B211}
(1988) 213; Nucl. Phys.{\bf B281} (1987)41.

\bibitem{GRONAU}
M. Gronau, D. Wyler, Phys. Lett. {\bf B265} (1991) 172.

\bibitem{GABBIANI2}
I.I.Bigi, F. Gabbiani, Nucl.Phys. {\bf B352} (1991) 309.

\bibitem{GAMMABS}
M. Voloshin, M. Shifman, N. Uraltsev, V. Khoze,
Sov. J. Nucl. Phys. {\bf 46} (1987) 112;
I. Bigi, B. Blok, M. Shifman, N. Uraltsev,
A. Vainshtein, in: `B Decays', 2nd edition, S. Stone (ed.),
World Scientific, pg. 132.

\bibitem{RRR}
P. Ramond, R. G. Roberts,
G. G. Ross, Nucl. Phys. {\bf B406} (1993) 19.

\bibitem{RABY}
G. Anderson et al., Phys. Rev. {\bf D49} (1994) 3660.

\bibitem{BUCCELLA}
For a first attempt see: F. Buccella et al.,
Phys. Lett. {\bf B302} (1993) 319.

\bibitem{E687}
P.L. Frabetti et al., preprint Fermilab-Pub-071-E (1994)

\bibitem{CLEOII}
M.S. Alam et al., preprint CLEO CONF 94-14 (1994)


\bibitem{YAMAMOTO}
I.I. Bigi, H. Yamamoto, preprint UND-HEP-94-BIG09.






\end{thebibliography}
\end{document}